# Extension of Transformational Machine Learning: Classification Problems

Adnan Mahmud[1], Oghenejokpeme Orhobor[2], and Ross D. King[1]
[1] Department of Chemical Engineering, University of Cambridge, CB3 0AS, UK
[2] National Institute of Agricultural Botany, CB3 0LE, UK

**Abstract**

This study explores the application and performance of Transformational Machine Learning (TML) in drug discovery. TML, a meta learning algorithm, excels in exploiting common attributes across various domains, thus developing composite models that outperform conventional models. The drug discovery process, which is complex and time-consuming, can benefit greatly from the enhanced prediction accuracy, improved interpretability and greater generalizability provided by TML. We explore the efficacy of different machine learning classifiers, where no individual classifier exhibits distinct superiority, leading to the consideration of ensemble classifiers such as the Random Forest.

Our findings show that TML outperforms base Machine Learning (ML) as the number of training datasets increases, due to its capacity to better approximate the correct hypothesis, overcome local optima, and expand the space of representable functions by combining separate classifiers' capabilities. However, this superiority is relative to the resampling methods applied, with Near Miss demonstrating poorer performance due to noisy data, overlapping classes, and nonlinear class boundaries. Conversely, Random Over Sampling (ROS) provides a more robust performance given its resistance to noise and outliers, improved class overlap management, and suitability for nonlinear class boundaries.

## 1 Introduction
### 1.1 What is Transformational Machine Learning?

Transformational Machine Learning (TML) is a meta-learning algorithm in the field of machine learning that exploits common attributes shared across different domains to develop composite models that outperform traditional base models (Olier et al., 2021). Traditional supervised machine learning (ML) systems learn from labelled data to predict the labels of unseen samples. When there are multiple related ML tasks, TML transforms the inherent features of these tasks into extrinsic features, offering a novel representation of the new task.

In traditional ML models, inherent properties or descriptors are used to construct predictive models. These models predict the probability that a particular input, such as an animal, belongs to a certain class. TML takes this a step further by using predictions from base models as extrinsic features for further learning. For example, a base model that predicts if a creature is a donkey based on intrinsic attributes like height and number of legs. This base model approach can be extended with the TML by considering properties extrinsic attributes shared by other animals, like donkeys and tigers. This allows TML to potentially incorporate features that weren't initially considered, thus enhancing its predictive capabilities.

## 1.2 Relevance of Transformational Machine Learning?

Drug discovery involves identifying and developing new therapeutic compounds that target specific biological processes or disease pathways. It's a complex, time consuming, and expensive procedure, but essential for meeting medical needs and improving patient outcomes (Dickson & Gagnon, 2004). Improving the drug discovery process can also reduce the time and cost of drug development, thereby benefiting patients, healthcare systems, and society at large.

TML outperforms standard ML methods when sufficient data is available (Olier et al., 2021), particularly in classification problems. It addresses the limitations of traditional Quantitative Structure-Activity Relationship (QSAR) models used in drug discovery, which are often chemically challenging to interpret. While ensemble learning models like bagging, boosting, and stacking offer different prediction outcomes, they may limit the predictability and interpretability (Verma & Mehta, 2017). TML, at the intersection of these methods, effectively mitigates the generalization problem, offering a balanced approach to improve prediction accuracy without compromising interpretability.

In drug discovery, TML provides three key advantages (Olier et al., 2021). Firstly, it enhances prediction accuracy by leveraging knowledge from related tasks and combining various models' predictions. Secondly, it improves interpretability, allowing researchers to understand better the underlying chemical relationships, leading to more informed decisions. Lastly, TML offers greater generalizability, providing more reliable predictions and feature selections for interpretation, thereby contributing to a more efficient drug discovery process.

## 1.3 Method Description

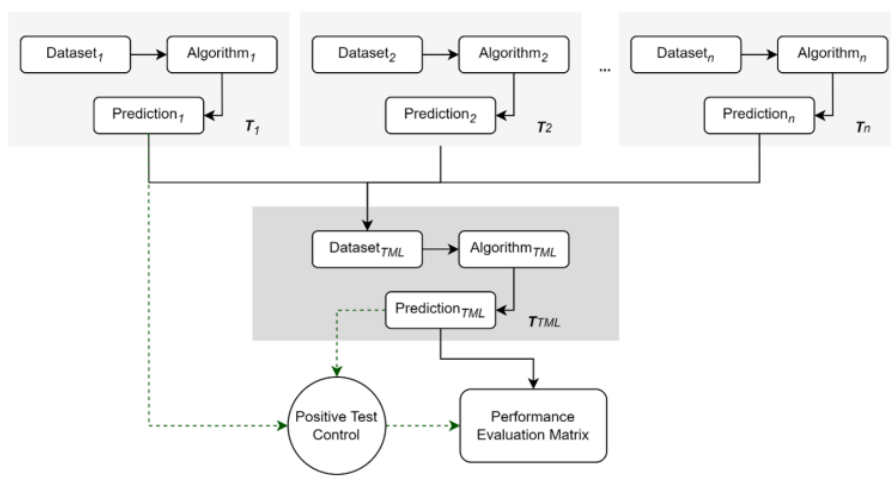

*Figure 1: Schematic representation of TML. The green dotted line represents the positive line represents the positive control test loop. The diagram is divided in three levels, where the top one being level I.*

Each learning task is illustrated as $T_i$. We focus on learning task $T_1$. In level I, baseline learning is used to learn a given algorithm (e.g., Random Forest) for all the tasks. In level II the predictions from the different learned base algorithm's predictions are used to learn a TML model for task $T_1$. In level III The performances of base ML and TML are evaluated. A positive control test was conducted by adding the "correct answer" in the TML prediction and observing higher performance compared to pure TML (detailed breakdown of positive control test in section 5).

## 2 Evaluation Step

The selection of evaluation metrics are particularly challenging in imbalanced classification scenarios due to two primary issues: most common metrics presuppose balanced class distribution, and not all prediction errors are equivalent across diverse classes in imbalanced classification.

Though accuracy is often considered a simple and intuitive metric, it can prove insufficient for imbalanced datasets, given its inability to account for class distribution (Juba & Le, 2019). This limitation can lead to misleading results under class imbalance. Consequently, precision, recall, and F1 score, metrics that consider all class performance, are frequently used in such situations.

In the following analysis, accuracy, precision, and recall metrics will be employed. Precision, defined as the proportion of true positives to the sum of positive predictions, is useful for minimizing false positives, an important aspect in fields like medical diagnosis or drug discovery where inaccurate positives can carry serious consequences.

Recall, also known as sensitivity or true positive rate, signifies the percentage of accurately detected positives among actual positives. This metric is particularly significant in scenarios where minimizing false negatives is crucial, for example, in diagnosing rare diseases or drug discovery. High recall suggests the model accurately identifies most positive instances.

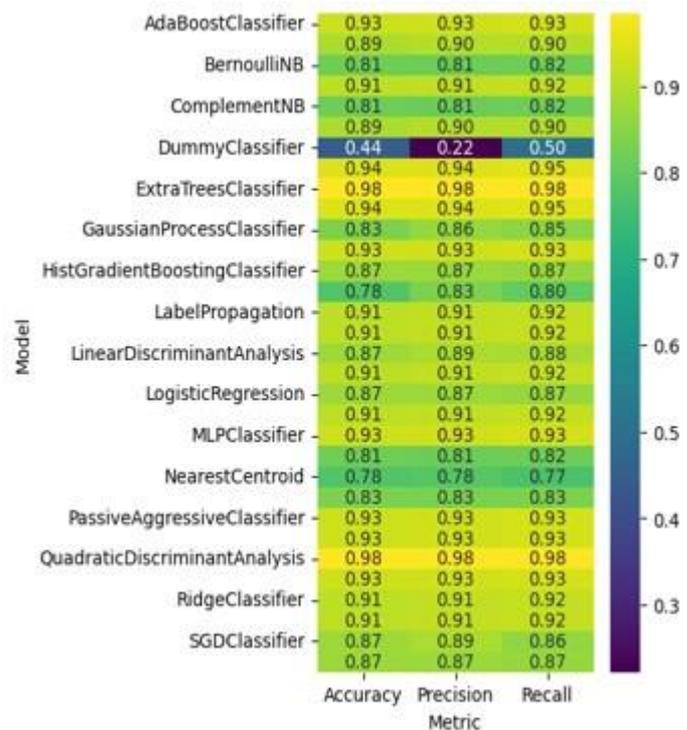

### 2.1 Benchmarking Algorithms

As demonstrated in *Figure 2*, no individual classifier demonstrates distinct superiority. In such contexts, an ensemble classifier such as the Random Forest often proves to be optimal. By merging results from multiple decision trees, it mitigates the impact of individual errors and biases, thereby engendering a model that is both robust and precise. The flexibility of Random Forests makes them apt for datasets insensitive to specific models, as they can judiciously utilize the strengths of various base classifiers to enhance overall performance. In contrast, the Gaussian NB, albeit simplistic, is computationally efficient and can exhibit high performance under certain conditions. Comparing Gaussian NB with Random Forest might provide valuable insights into whether an interpretable model such as Gaussian NB suffices, or if the added complexity of an ensemble approach is needed for improved outcomes.

*Figure 2: Benchmarking Classifier Performance: Heatmap of Accuracy, Precision,*

## 3 Pseudo Code

---
**Algorithm 1** Transformational Machine Learning
---
1: **Step 1:** Import necessary libraries
2: import necessary libraries
3: **Step 2:** Define classifiers
4: classifiers = define classifiers()
5: **Step 3:** Define resampling methods
6: resampling methods = define resampling methods()
7: **Step 4:** Main loop
8: **for** classifier in classifiers do
9:   **for** resampling method in resampling methods do
10:     **Step 4.1:** Perform data pre-processing
11:     dataset = load dataset()
12:     train data, test data = split dataset(dataset, cutoff values)
13:     resampled data = resample dataset(train data, resampling method)
14:     save processed datasets(resampled data)
15:     **Step 4.2:** Train the base models
16:     processed datasets = load processed datasets()
17:     base model = train base model(classifier, processed datasets)
18:     save trained base model(base model)
19:     base model results = evaluate base model(base model)
20:     save results(base model results)
21:     **Step 4.3:** Create the TML input datasets
22:     base models = load base models()
23:     tml input datasets = generate TML input datasets(base models)
24:     save TML input datasets(tml input datasets)
25:     **Step 4.4:** Train and evaluate the TML models
26:     TML input datasets = load TML input datasets()
27:     TML model = train TML model(base model, TML input datasets)
28:     TML model results = evaluate TML model(TML model)
29:     save results(TML model results)
30: **Step 5:** Combine results and save them in a CSV file
31: combined results = combine results()
32: save results to csv(combined results)

---

## 4 Results & Discussion

Similar to the regression analysis (Olier et al., 2021), Transformational ML (TML) outperforms the base ML as the number of training datasets increases. This observation can is owing to a combination of the three factors mentioned by Dietterich (2000):

**I. Statistical:** the ensemble technique is better able to approximate the actual hypothesis as the amount of training data rises by "averaging" the votes of several accurate classifiers. This decreases the possibility of selecting the incorrect classifier, as the ensemble is more likely to settle on the correct hypothesis. The bigger training set enables the learning algorithm to generate more precise hypotheses inside the hypothesis space.

**II. Computational:** with additional training data, the ensemble technique may efficiently explore the search space by executing local search algorithms from a variety of starting locations. This improves the possibility of local optima being surpassed and global optima being attained. As a result of aggregating the output of numerous classifiers, each of which may have discovered various local optima, the ensemble delivers a more accurate estimate of the genuine unknown function.

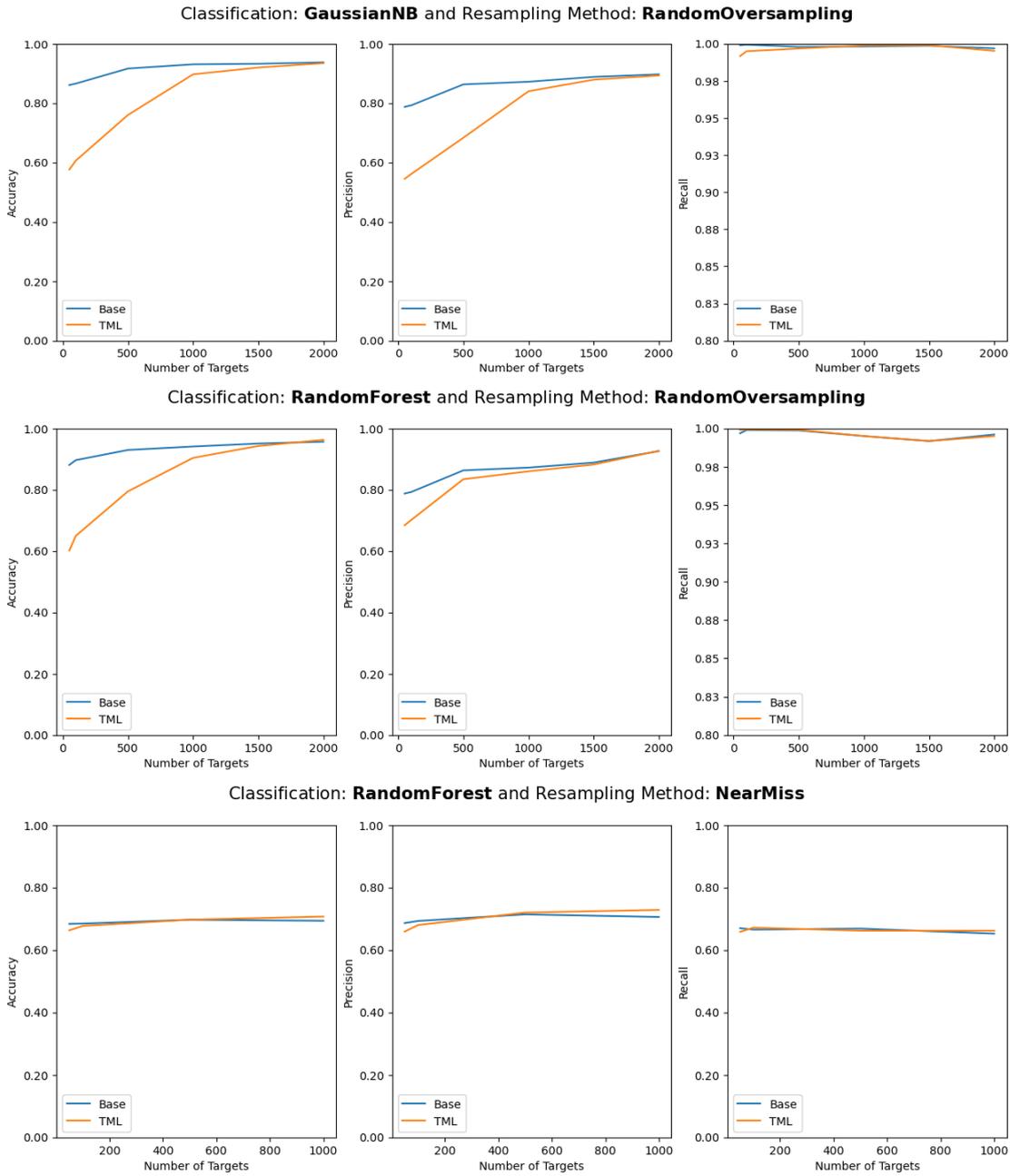

*Figure 3: Comparative Analysis of (top to bottom): (i) **Random Forest Classifier** performance using **Random Oversampling** method, (ii) **Gaussian Naïve Bayes Classifier** performance using **Random Oversampling** method, (iii) **Random Forest Classifier** performance using **Near Miss Undersampling** method. This graph illustrates the relationship between the number of targets (X-axis) and the performance metrics Accuracy, Precision, and Recall (Y-axis).*

**III. Representational:** when the true function cannot be represented by a single hypothesis in the hypothesis space, the ensemble can expand the space of representable functions by creating weighted sums of hypotheses. With additional training data, the ensemble technique can discover more diverse and precise weighted combinations of hypotheses, which may approximate the real function more closely.

Therefore, the TML outperforms the basic ML as the number of training datasets rises due to its capacity to better approximate the correct hypothesis, overcome local optima, and expand the space of representable functions by combining the capabilities of separate classifiers.

For resampling methods like Near Miss, the fact that TML performs nearly as well as Base ML is because Base ML performs poorly in all metrics to begin with, especially compared with Random Over Sampling. This implies that it requires relatively less target for the TML to outperform Base ML. As seen in *Figure 3*, it takes around 1000 targets for TML to match Base ML performance.

### 4.1 Why Recall is usually higher for both Base ML and TML?

In cases where the class distribution is imbalanced and the positive class (i.e., the class of interest) is uncommon compared to the negative class, a model may have lower accuracy but higher recall (He and Garcia, 2009). In such situations, a model that always predicts the negative class will have high accuracy but low recall, as it will miss many instances of the positive class. In contrast, a model that predicts the positive class more frequently, even if it makes more mistakes, will have a higher recall because it captures a greater proportion of positive instances (He and Garcia, 2009).

Similarly, a model can have lower accuracy but greater precision if it excels at identifying the positive class but has difficulty with the negative class (Davis and Goadrich, 2006). In such situations, the model may have a high precision because it makes few false positive predictions (i.e., it correctly identifies positive instances), but it also makes many false negative predictions (i.e., it incorrectly identifies many negative instances), resulting in a lower accuracy (Davis and Goadrich, 2006).

### 4.2 Possible Reasons for Base ML to perform poorly with Near Miss for a QSAR classification analysis?

There are circumstances in which the use of Near Miss may not be optimal.

**I. Noisy data:** The applicability of Near Miss to the QSAR dataset may be compromised if the data is excessively noisy or contains outliers (Batista, Prati, and Monard, 2004). The method of selecting instances of the majority class based on their proximity to instances of the minority class may result in the retention of incoherent samples or outliers that are in close proximity to instances of the minority class. This may have an adverse effect on the model's efficacy (Batista, Prati, and Monard, 2004).

**II. Highly overlapping classes:** Classes with a lot of overlap in cases where there is significant overlap between the classes (class 0 and class 1) in the dataset, Near Miss may be less effective (He and Ma, 2013). This occurs when minority and majority class instances are intermixed in the feature space. This is because the elimination of majority class instances in close proximity to minority class instances may result in the loss of valuable discriminatory information between the classes (He and Ma, 2013).

**III. Nonlinear class boundaries:** In instances where classes are separated by nonlinear boundaries, Near Miss may be rendered ineffective (He and Ma, 2013). By eliminating instances from the majority class based on their proximity, the method assumes a simple and linear decision boundary. The aforementioned occurrence has the potential to result in distorted decision boundaries and a subsequent decrease in the model's performance (He and Ma, 2013).

### 4.3 Why is Random Over Sampling (ROS) preferable?

Under the provided conditions, ROS may be a preferable alternative to the QSAR dataset due to the following factors.

**I. Resistance to noise and outliers:** ROS generates new synthetic instances of the minority class by arbitrarily sampling with replacement from the initial instances of the minority class (Chawla, Bowyer, Hall, and Kegelmeyer, 2002). ROS is less susceptible to noise and anomalies in the dataset because it emphasises increasing the representation of the minority class as opposed to decreasing the representation of the dominant class (Chawla, Bowyer, Hall, and Kegelmeyer, 2002).

**II. Improved administration of class overlap:** ROS does not directly affect the preponderance of class instances (Fernández, García, Herrera, and Chawla, 2018). ROS may aid in class distribution balance by generating additional instances of the minority class without removing potentially advantageous majority class instances that are adjacent to minority class instances. This allows the classifier to understand the class boundaries more effectively in situations where classes overlap substantially (Fernández, García, Herrera, and Chawla, 2018).

**III. Suitable for nonlinear class boundaries:** ROS does not assume a particular decision boundary shape across classes, making it appropriate for nonlinear class boundaries (Fernández, García, Herrera, and Chawla, 2018). By increasing the representation of the minority class, the classifier can more effectively understand the relationship between characteristics and classes, even when the class boundaries are nonlinear. This can result in improved model performance compared to strategies such as Near Miss, which rely on assumptions regarding the decision boundary (Fernández, García, Herrera, and Chawla, 2018).

## 5 Positive Control Test

A positive control is a subset of a control group and is defined as an experiment that undergoes treatment with a known outcome and, as such, should demonstrate a certain change throughout the experiment.

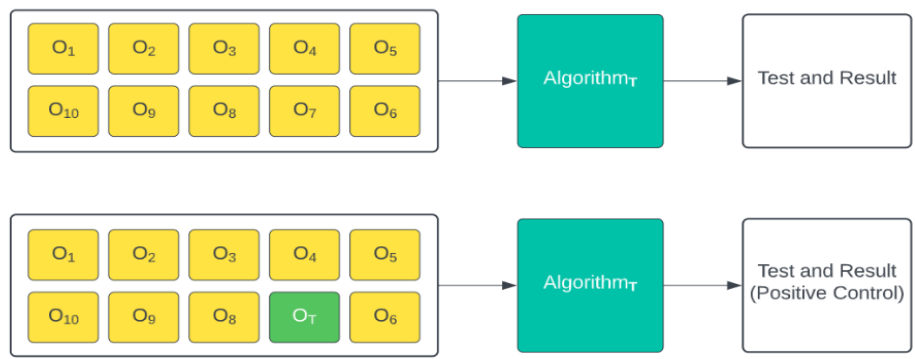

*Figure 3: Schematic demonstration of Positive Control in TML*

Let the output from 10 arbitrary training algorithms, from a set of given input datasets, be $O_n$ (represented as yellow rectangles), where $1 \leq n \leq 10$. Let the dataset of interest in this particular example and its training algorithm be defined as $O_T$ (represented as a green rectangle) and $Algorithm_T$.

In a non-controlled TML experiment, $Algorithm_T$ will be processed with a set of output datasets, $O_n$ (in the above example 10 datasets), trained by their respective training algorithms. Thereafter, as the number of *n* increases the performance of TML gets better compared to the baseline ML methods. However, in a *positive controlled* TML experiment, one of the $O_n$ is replaced with $O_T$, which colloquially can be termed as "hiding the right answer in the mix". Given that the underlying codes for the TML are in the correct order, the $Algorithm_T$ should perform better than the un-controlled case, since the right set of answers was embedded in the datasets.

*Table 1: Results of positive control in a tabular format. The italic numbers represent the highest scoring values in the given criteria/column.*

|  | **Accuracy** | **Precision** | **Recall** |
|---|---|---|---|
| **Non-Controlled TML** | 0.685 | 0.674 | 0.713 |
| **Positive Controlled TML** | *0.917* | *0.924* | *0.910* |
| **Base ML** | 0.897 | 0.903 | 0.890 |

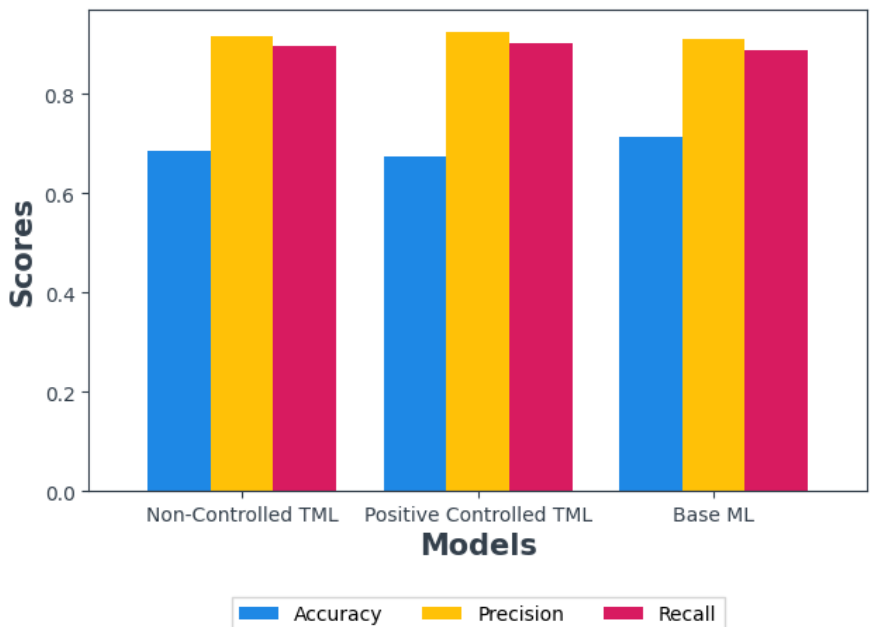

*Figure 4: Comparison between non-controlled TML, positively controlled TML and Base ML method.*

From *Table 1* and *Figure 5*, it can be deduced that positively controlled TML has performed better than the non-controlled TML experiment and baseline ML method. Therefore, this indicates that the underlying codes have a higher chance of being correct.

## 6 Statistical Testing

The Wilcoxon signed-rank (Woolson, 2008) test was used to compare the Traditional Machine Learning (TML) method and the base methods due to its compatibility with real-world datasets like QSAR, its suitability for comparing related samples, its applicability to ordinal or continuous data, and its resilience to outliers. However, it's important to note its limitations, such as lower power when parametric test assumptions are satisfied, inability to quantify the degree of difference, and less reliable results with small samples or numerous ties.

*Table 2: List of Null Hypothesis and Alternative Hypothesis for all three Wilcoxon Single Rank tests on accuracy, precision and recall.*

| **Test** | **Null Hypothesis** | **Alternative Hypothesis** |
|---|---|---|
| Accuracy | As the number of datasets increases, TML accuracy significantly does not outperform Base accuracy. | As the number of the dataset increases, TML accuracy significantly outperforms Base accuracy. |
| Precision | As the number of datasets increases, TML precision significantly does not outperform Base recall. | As the number of the dataset increases, TML precision significantly outperforms Base precision. |

| Recall | As the number of datasets increases, TML recall significantly does not outperform Base recall. | As the number of the dataset increases, TML recall significantly outperforms Base recall. |

The results indicated that TML significantly outperforms the base methods in terms of both accuracy and precision, with the null hypotheses for these measures being rejected at a P-value of *0.031*. This suggests TML could potentially be advantageous in QSAR dataset applications for drug development. However, the study found no significant difference in recall between the TML and base methods, accepting the null hypothesis with a P-value of *0.156*. Consequently, the TML method's capacity to accurately identify positive cases was not demonstrably superior. These findings underline the need for further research to optimize TML's performance, particularly regarding recall.

## 7 Conclusion

This study demonstrates the potential of using Transformational Machine Learning (TML) to improve predictions in the complex process of drug discovery. The results show that TML models outperform conventional Machine Learning models as more training data becomes available. This highlights TML's ability to learn generalized patterns across domains and develop robust composite models. Near Miss resampling was found to be less effective for TML due to issues with noisy data and class overlap. However, Random Over Sampling proved to be a good complement to handle noise, imbalance, and nonlinearity. Overall, this research illustrates that the meta-learning capabilities of TML can enhance accuracy, interpretability, and generalizability compared to individual classifiers. Drug discovery is an promising application for TML where leveraging insights across multiple datasets is advantageous. Further research should explore optimal TML architectures and resampling techniques to maximize performance. But these initial findings indicate that TML has considerable potential to expedite the drug discovery process through improved predictive modelling.

## Reference

Batista, G.E., Prati, R.C. and Monard, M.C., 2004. A study of the behavior of several methods for balancing machine learning training data. SIGKDD explorations, 6(1), pp.20-29.

Chawla, N. V., Bowyer, K. W., Hall, L. O., & Kegelmeyer, W. P. (2002). SMOTE: synthetic minority over-sampling technique. Journal of artificial intelligence research, 16, pp.321-357.

Davis, J. and Goadrich, M., 2006. The relationship between Precision-Recall and ROC curves. In Proceedings of the 23rd international conference on Machine learning, pp.233-240. ACM.

Dickson, M. and Gagnon, J.P. (2004) 'Key factors in the rising cost of New Drug Discovery and Development', Nature Reviews Drug Discovery, 3(5), pp. 417–429. doi:10.1038/nrd1382.

Dietterich, T.G. (2000). Ensemble Methods in Machine Learning. In: Multiple Classifier Systems. MCS 2000. Lecture Notes in Computer Science, vol 1857. Springer, Berlin, Heidelberg.

Fernández, A., García, S., Herrera, F., & Chawla, N. V. (2018). SMOTE for learning from imbalanced data: progress and challenges, marking the 15-year anniversary. Journal of artificial intelligence research, 61, pp.863-905.

He, H. and Garcia, E.A., 2009. Learning from imbalanced data. IEEE Transactions on knowledge and data engineering, 21(9), pp.1263-1284.

He, H. and Ma, Y., 2013. Imbalanced learning: foundations, algorithms, and applications. John Wiley & Sons.

Juba, B. and Le, H.S. (2019) 'Precision-recall versus accuracy and the role of large data sets', Proceedings of the AAAI Conference on Artificial Intelligence, 33(01), pp. 4039–4048. doi:10.1609/aaai.v33i01.33014039.

Olier, I. et al. (2021) "Transformational machine learning: Learning how to learn from many related scientific problems," Proceedings of the National Academy of Sciences, 118(49). Available at: https://doi.org/10.1073/pnas.2108013118.

Verma, A. and Mehta, S. (2017) 'A comparative study of Ensemble Learning Methods for classification in bioinformatics', 2017 7th International Conference on Cloud Computing, Data Science &amp; Engineering - Confluence [Preprint]. doi:10.1109/confluence.2017.7943141.

Woolson, R.F. (2008) "Wilcoxon signed-rank test," Wiley Encyclopedia of Clinical Trials. Available at: https://doi.org/10.1002/9780471462422.eoct979.